\documentclass[11pt]{article}

\usepackage{fancybox}
\usepackage[dvips]{graphicx}
\usepackage{array}
\usepackage{lscape}

\evensidemargin=\oddsidemargin
\begin{document}

\title{An Analysis of Transaction and Joint-patent Application Networks}
\author{Hiroyasu Inoue\\
Osaka Sangyo University, Osaka 574-0013, Japan}
\date{}

\maketitle


\begin{center}
\begin{minipage}[h]{14.0cm}
\begin{center}{\bf Abstract}\end{center}
Many firms these days,
forced by increasing international competition and
an unstable economy, are opting to
specialize rather than generalize as a way of maintaining their competitiveness.
Consequently,
they cannot rely solely on themselves, but must
cooperate by combining their advantages.

To obtain the actual condition for this cooperation,
a multi-layered network
based on
two different types of data was investigated.
The first type was transaction data from Japanese firms.
The network created from the data included 961,363 firms and
7,808,760 links.
The second type of data were from joint-patent applications in Japan.
The joint-patent application network included
54,197 nodes and
154,205 links.
These two networks were merged into one network.

The first anaysis was based on input-output tables
and three different tables were compared.
The correlation coefficients between tables
revealed that
transactions were more strongly tied to joint-patent applications
than the total amount of money.
The total amount of money and transactions
have few relationships
and these are probably connected to joint-patent applications
in different mechanisms.
The second analysis was conducted based on 
the $\mbox{p}^{*}$ model.
Choice, multiplicity, reciprocity, multi-reciprocity and
transitivity configurations were evaluated.
Multiplicity and reciprocity configurations were
significant in all the analyzed industries.
The results for multiplicity 
meant that transactions and joint-patent application links 
were closely related.
Multi-reciprocity and transitivity configurations were
significant in some of the analyzed industries.
It was difficult to find any common
characteristics in the industries.
Bayesian networks were used in
the third analysis.
The learned structure revealed that
if a transaction link between two firms is known,
the categories of firms' industries do not affect to the existence of a patent link.
\end{minipage}
\end{center}

\section{Introduction}
\label{cha:int}

Many firms these days,
forced by increasing international competition and
an unstable economy, 
are opting to
specialize rather than generalize as a way of maintaining their competitiveness.
Consequently,
they cannot rely solely on themselves, but must
cooperate by combining their advantages.
Although there are many ways for them to do this,
in terms of competitiveness,
the most important objective is
creating novel products and services.

Intuitively,
if firms create something new,
they can sell more of their goods or services than they used to.
The priority of this process is that cooperative research and
development (R\&D) comes first
and transactions come second.
However, firms cannot know what is innovative for their customers
without selling any goods or services to them.
Hence, it is natural
for sellers to try to know what buyers want as much as possible.
Consequently,
the relationship between sellers and buyers
develops and yields a new relationship as cooperative R\&D.
The priority in this process is that transactions come first
and cooperative R\&D comes second.
Therefore, if we focus on the relationships between
cooperative R\&D and transactions,
it is difficult to determine which relationship is built first.
Since firms have to strategically cooperate more than ever,
pursuing this question has become increasingly important.
This question can be rephrased as ``seeds and needs,'' i.e.,
do seeds precede needs, or do needs precede seeds?


Based on the above,
this paper investigates
the relationship between cooperative R\&D
and transactions.
The author acquired data that included exhaustive transactions
and joint-patent applications made by Japanese firms.
Based on the data, this paper discusses three different analyses.

The first is based on input-output tables.
Input-output tables and their analyses
were first developed by Leontief \cite{Leontief86} and 
he used a matrix representation
of the economy of some actors to predict
the effect of changes
in one industry on others
and by others (consumers, government, and foreign suppliers.)
However, 
the effect by others has not been considered in this paper.
The input-output tables are used to
analyze economics, and more concretely, the total amount of money
flowing between industries.
The present author created other types of matrices
based on transactions
(number of relationships between firms, not the total amount of money)
and cooperative R\&D.
By comparing matrices,
he discusses
the relevance of cooperative R\&D and other economic activities.
The most important point is that this analysis
is conducted on industries that are aggregates of firms
and this is a basis for the second analysis.

The second analysis is conducted based on configurations of networks
whose nodes are firms and links are transactions and cooperative R\&D.
Actors in matrices are industries
and the relationships are investigated in the first analysis.
However, 
actors are firms in the second analysis
and more complicated structures between them are investigated.
In the analysis, the present author
uses the $\mbox{p}^{*}$ model \cite{Wasserman05} (this has recently been called
the exponential random graph (ERG) model.)
This method reveals whether a network
includes significant configurations (some specific patterns or motifs) or not.

Bayesian networks are used in the third analysis.
The Bayesian networks are completely different from the networks
mentioned in this paper thus far and they are methods of
finding causality from data.
The Bayesian networks were first proposed by
Pearl \cite{Pearl09}.
One of the objectives of this paper is
to investigate which link (transactions or cooperative R\&D)
is created first.
This means that we have to establish the causality between relationships.
If one tends to be created first,
that is evidence of causality.

The rest of this paper is organized as follows.
Section 2 presents
the data used in this paper.
Section 3 explains the preparation of the input-output tables as the analysis.
Section 4 explains the $\mbox{p}^{*}$ model for the second analysis,
and Section 5 explains Bayesian networks for the third analysis.
Section 6 concludes this paper.


\section{Data}

\begin{figure}[b]
\begin{center}
\includegraphics[scale=0.4]{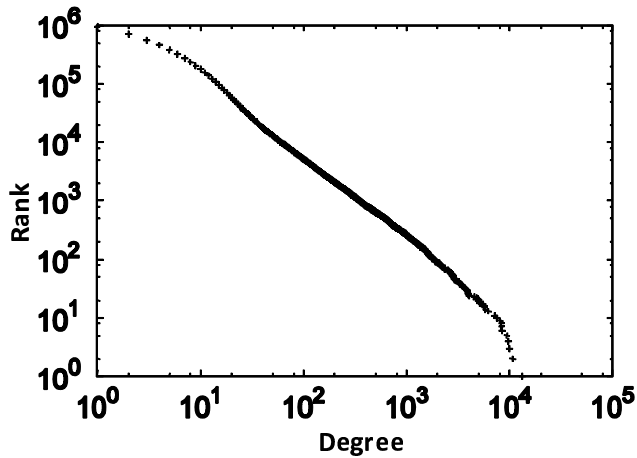}
\caption{Degree distribution by rank in transaction network.
Horizontal axis indicates
degree and vertical axis indicates
rank counted from highest degree.}
\label{fig:degreeTrans}
\end{center}
\end{figure}

This paper explains the two different types of data that were used.
The first type was transaction data from Japanese firms
provided by Tokyo Shoko Research (a Japanese investigation firm).
The data included 961,363 firms and
7,808,760 transactions that took place between them in 2005.
The current author created a network from the data and
the it had firms as nodes,
and transactions as links.
A link had a direction and
a direction meant the flow of money.
There is a degree distribution in Figure \ref{fig:degreeTrans}
that explains the whole structural characteristics.
The horizontal axis indicates
the degree and the vertical axis indicates
the rank counted from the highest degree.
We can see that
the graph can be roughly fitted with a line
except for the low degree; therefore,
it does not have a peak like p normal distribution.

\begin{figure}[b]
\begin{center}
\includegraphics[scale=0.4]{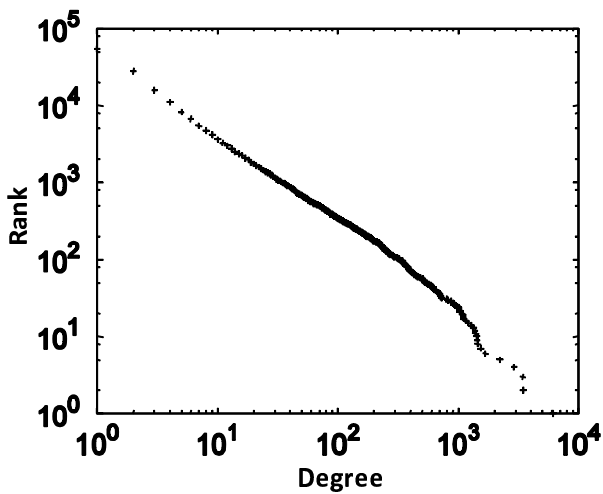}
\caption{Degree distribution by rank in joint-patent application network.
Horizontal axis indicates
degree and the vertical axis indicates
rank counted from highest degree.}
\label{fig:degreePatent}
\end{center}
\end{figure}

The other data are joint-patent applications.
These data were provided by Koukai Tokkyo Kouhou
(publication issued by the Japanese Patent Office.)
The data included 5,570,786 patents.
The objective of this paper
is to investigate the relationships between transactions and cooperative R\&D
and these patent data are used to identify cooperative R\&D activity.
Although cooperative R\&D involves a wide variety of activities,
patents are one of the most decisive items of evidence for cooperative R\&D.
A patent mainly has information on new findings
and also has other information
about the people or organizations that published the patent.
This paper uses applicants that are in charge of patent applications.
Most of the applicants are firms and
this paper only treats firms.
Therefore, applicants that are firms need to be distinguished from other types of
applicants.
As firms have a coorporate status in their names, this can be used to
distinguish them.
A patent possibly has more than one firm that are applicants,
and that can be regarded as evidence of cooperative R\&D activity.
Hence, the network for joint-patent applications
is composed of applicants as nodes,
and joint-patent applications as links.
If more than two nodes are involved in a patent,
those nodes are connected in the manner of a complete graph.
The joint-patent application network includes
54,197 nodes and
154,205 links between the firms.
The links do not have a direction.
The degree distribution is plotted in Figure \ref{fig:degreePatent}
in the same way as in Figure \ref{fig:degreeTrans}.
The horizontal axis indicates the
degree and the vertical axis indicates 
the rank counted from the highest degree.

Inoue et al. recently studied the joint-patent application network \cite{Inoue10}.
Their study included an investigation into the existence
of a power law in the degree distribution of the network.
The least-squares method is generally
used for verification,
and the inclination (the power index) in the power-law distribution is estimated.
However, this approach has two critical flaws.
The first is that people cannot determine
whether the distribution really follows a power law or not.
The inclination can be estimated
for any kind of distribution
with this method.
The second flaw is that
people cannot determine which part of the distribution follows a power law.
Consequently,
we have the method created by Clauset, et al. \cite{Clauset09}
that combines maximum-likelihood fitting methods
with goodness-of-fit tests based on the Kolmogorov-Smirnov statistic.
If the probabilistic distribution of degree
follows a power law,
the equation for a discrete case can be expressed by
\[
p(x)=\frac{x^{-\alpha}}{\zeta(\alpha,x_{\mbox{min}})},
\]
where
\[
\zeta(\alpha,x_{\mbox{min}})=\sum^{\infty}_{n=0}(n+x_{\mbox{min}})^{-\alpha},
\]
Here, $x_{\mbox{min}}$ is the lower bound
and $\alpha$ is the scaling parameter.
$x_{\mbox{min}}$ is necessary because
$p(x)$ diverges at $x=0$.
The equation 
for the cumulative distribution
is defined as
\[
P(x)=\frac{\zeta(\alpha,x)}{\zeta(\alpha,x_{\mbox{min}})}.\label{eqn:c}
\]
As a result,
the network for joint-patent applications is scale-free and
$\alpha$ is 2.03 at $x_{\mbox{min}}=7$.

The two networks for transactions and joint-patent applications
are different.
However, these two networks have firms and 
they can therefore be merged into one network
that has firms as nodes,
and transactions and joint-patent applications as links.
These links should be distinguished;
hence, the network includes two different types of links.
The network has 975,607 nodes and
this number is less than the sum of the number of nodes in
the networks.
This is because there are identical firms.
The names and addresses of firms are used to identify them.
Many nodes are duplicated
if only names are used.
If both are used,
no duplication occurs and each node can be uniquely distinguished.
The network is called a multi-layered network in this paper.

\section{Input-output tables}
\label{cha:iotable}

The first analysis is based on input-output tables
that were developed by Leontief \cite{Leontief86}.
An industry buys primary material or fuel,
fabricates products from them,
and sells these to other industries.
Input-output tables list
how many goods and services are produced and sold between industries.
These tables are expressed by matrices.
These matrices are very useful for calculating
how much production increases in industries
when some industry has exceptional needs.
Japanese input-output tables are well known
for their accuracy.
The original input-output table 
included non-industrial actors
(consumers, government, and foreign suppliers),
but these have been omitted from this paper.

This paper's objective is to discuss the relationships
in transactions and joint-patent applications between firms.
Input-output tables, on the other hand, handle the relationships between industries.
Needless to say, industries are different from firms
and they are aggregates of firms.
The objective of this section is to explain
the macroscopic trend in the relationships between firms.

Table \ref{tbl:iosample} is a pedagogic example of an input-output table.
This table is for
``electrical machinery'' and ``information and communication machinery'' industries.
The rows mean output industries and
the columns mean input industries.
Therefore, 354 billion Japanese yen passed from the electrical machinery industry
to the information and communication machinery industry.
As we can see from the table
the original input-output table
provides flows for the total amount of money between industries.
The matrix for the table was called $\mathbf{M_a}$.

The data for the input-output tables
were published by the Japanese Statistics
Bureau, and the Director-General for the Policy Planning and Statistical Research
and Training Institute.
There are many input-output tables and
this paper uses tables from major consolidated industries
\footnote{In the English manuscript, there are 32 major consolidated industries.
However, the one in Japanese has 34.
This paper uses a table divided into 34 industries.}.
The number of major consolidated industries is 34.
All of the industries are shown in Table \ref{tbl:pstarresult1} and \ref{tbl:pstarresult2}.

\begin{figure}[b]
\begin{center}
\includegraphics[scale=0.5]{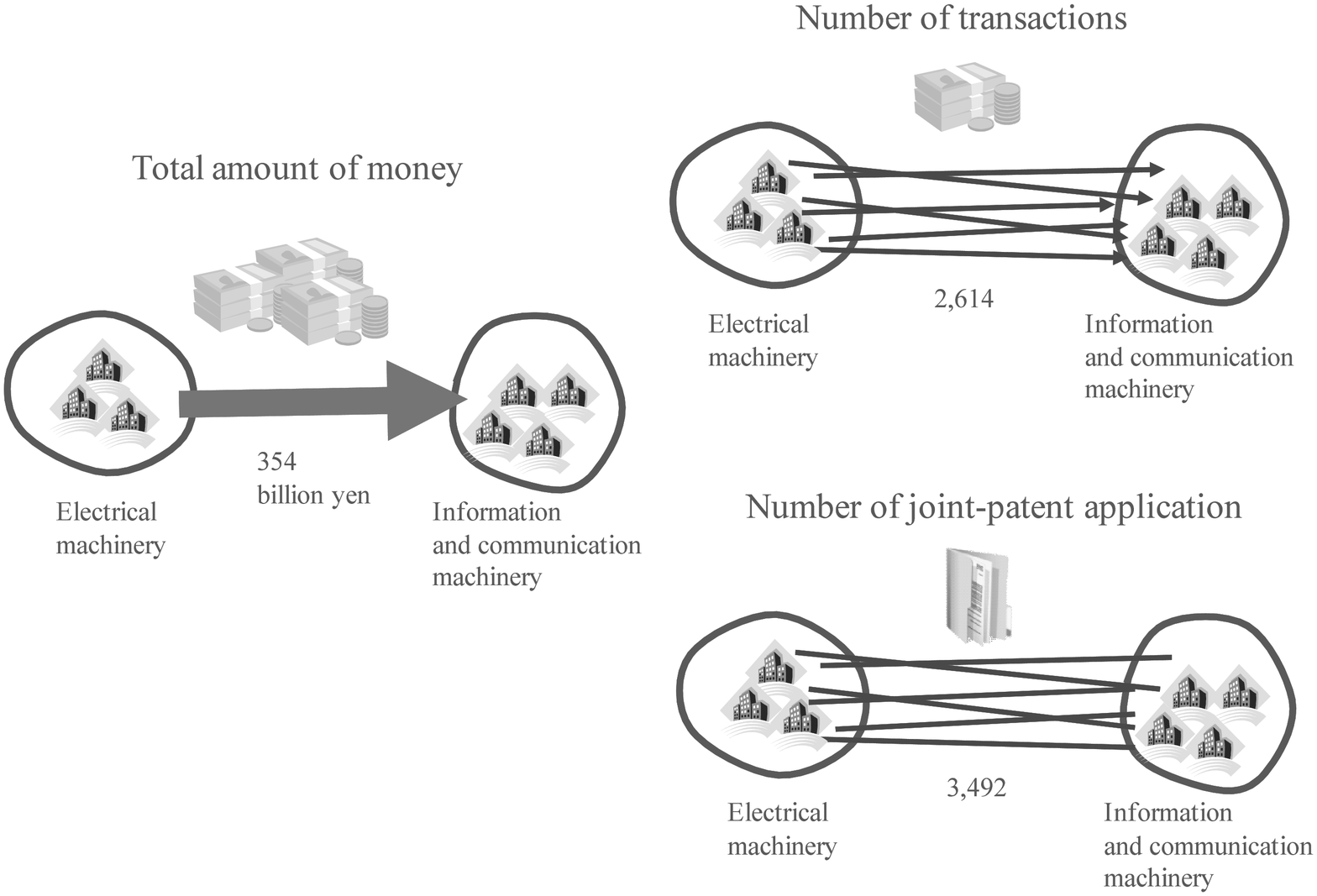}
\caption{Difference between three matrices. These figures are examples and indicate how
value of each matrix element is determined.}
\label{fig:matMerge}
\end{center}
\end{figure}

However, we can create the matrices
based on the number of transactions
between firms, not the total amount of money
or the number of joint-patent applications.
By comparing three different matrices
we can discuss the relevance between them.
The author made two new 34 times 34 matrices.
The matrix for the transation table is called $\mathbf{M_t}$,
and the joint-patent application table is called $\mathbf{M_p}$.
Figure \ref{fig:matMerge} shows the difference between three matrices.
The schematic at left is
for a situation where money is flowing from an electrical machinery industry
to an information and communication machinery industry ($\mathbf{M_a}$).
Three hundred fifty four billion yen was assigned to
the element of the matrix that corresponds to this pair.
The upper right schematic is for a situation
situation where money is flowing from an electrical machinery industry
to an information and communication machinery industry.
However, the value of the matrix is the number of transactions
conducted by firms between industries ($\mathbf{M_t}$).
The lower right figure is for a
situation where firms have jointly applied for patents
in an electrical machinery industry
and those in an information and communication machinery industry ($\mathbf{M_p}$).

\begin{figure}[b]
\begin{center}
\includegraphics[scale=0.35]{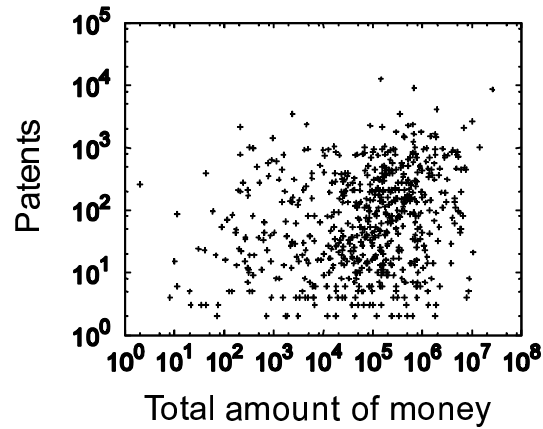}
\caption{Scatter plot for $\mathbf{M_a}$ and $\mathbf{M_p}$.
Each plot corresponds to elements
that have same position in $\mathbf{M_a}$ and $\mathbf{M_p}$.
Correlation coefficient is 0.31.}
\label{fig:massPatent}
\end{center}
\end{figure}

\begin{figure}[b]
\begin{center}
\includegraphics[scale=0.35]{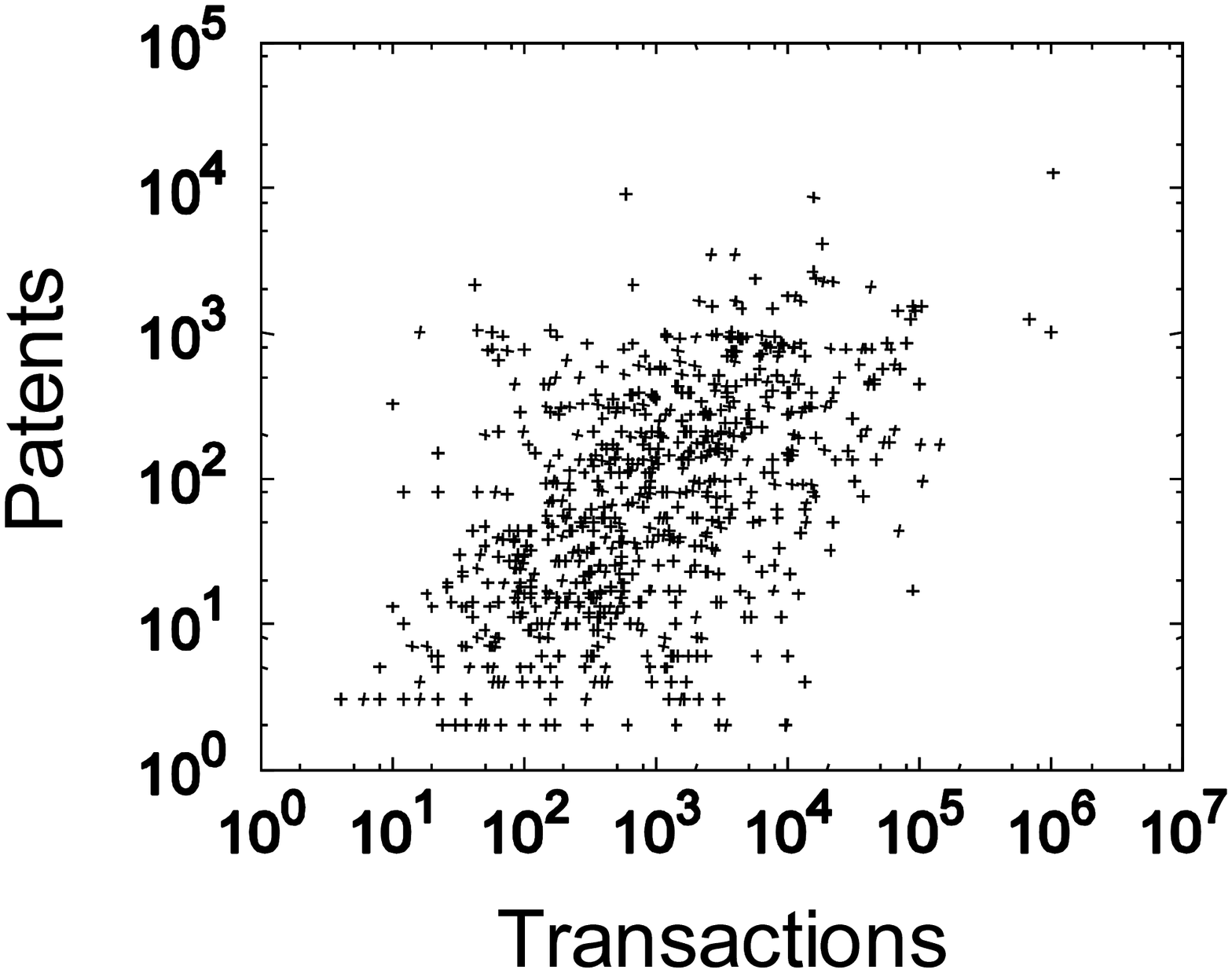}
\caption{Scatter plot for $\mathbf{M_t}$ and $\mathbf{M_p}$.
Correlation coefficient is 0.45.}
\label{fig:transPatent}
\end{center}
\end{figure}

\begin{figure}[b]
\begin{center}
\includegraphics[scale=0.35]{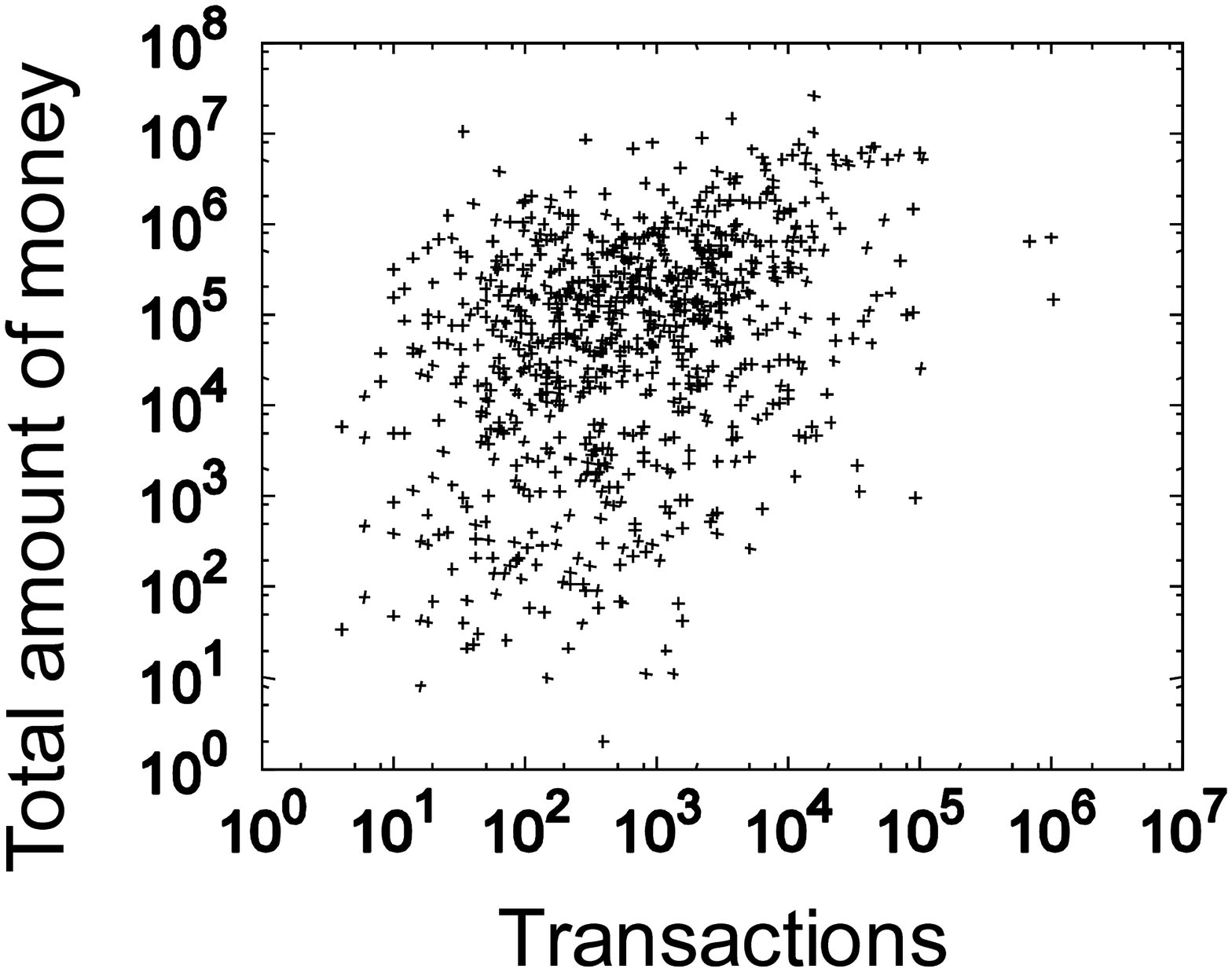}
\caption{Scatter plot for $\mathbf{M_t}$ and $\mathbf{M_a}$.
Correlation coefficient is 0.05.}
\label{fig:transMass}
\end{center}
\end{figure}

Figure \ref{fig:massPatent} is a scatter plot for $\mathbf{M_a}$ and $\mathbf{M_p}$.
Let $x$ represents the value for the horizontal axis
and $y$ be that for the vertical axis.
Then, each position on the plots corresponds to
\[
(x,y)=(\mathbf{M_a}_{ij},\mathbf{M_p}_{ij}).
\]
This means a plot corresponds to elements
that have the same position in $\mathbf{M_a}$ and $\mathbf{M_p}$.
Figure \ref{fig:transPatent} is a scatter plot for $\mathbf{M_t}$ and $\mathbf{M_p}$
and 
Figure \ref{fig:transMass} is a scatter plot for $\mathbf{M_t}$ and $\mathbf{M_a}$.

\begin{table}[t]
\begin{center}
\caption{Pedagogic example of input-output tables. Numbers are flows for total amount of money.
Money flows from industries in rows to ones in columns.}
\label{tbl:iosample}
\begin{tabular}{|l|rr|}
\hline
& \multicolumn{1}{|l}{Electrical machinery} & \multicolumn{1}{l|}{Information and communication} \\
& \multicolumn{1}{|l}{} & \multicolumn{1}{l|}{machinery} \\
\hline
Electrical machinery & 1,938 billion yen & 354 billion yen \\
Information and communication machinery & 2 billion yen & 422 billion yen \\
\hline
\end{tabular}
\end{center}
\end{table}

These figures are similar.
However, the Pearson's correlation coefficients for the scatter plots are
0.31 ($\mathbf{M_a}$ and $\mathbf{M_p}$),
0.45 ($\mathbf{M_t}$ and $\mathbf{M_p}$)
and 
0.05 ($\mathbf{M_t}$ and $\mathbf{M_a}$).
The objective of this paper is to discuss
the relationships in transactions and joint-patent applications between firms
and the results
reveal that 
either the
total amount of money or transactions
are strongly related to the number of joint-patent applications.
The correlation coefficients indicate
transactions are more strongly tied to joint-patent applications
than the total amount of money.
Since joint-patent applications represent the intensity
of scientific and technological connections between industries,
this means
that the total amount of money can
be used to track
scientific and technological connections to some degree,
but transactions are 
more preferable to reach the objective.
The correlation coefficient for $\mathbf{M_t}$ and $\mathbf{M_a}$
is important and
0.05 means the total amount of money and transactions
have few relationships
and these are probably connected to joint-patent applications
in different mechanisms.

\section{$\mbox{P}^{*}$ model}
\label{cha:pstar}

The previous section explained that
transactions are probably a better approach
to understanding the relationships to joint-patent applications than
the total amount of money
at the industry level.
This section
presents the results and discusses
configurations for the networks.
The number of
transactions and joint-patent applications
were counted for industries
in the previous section
but this section 
returns the units to firms.

In the analysis, the current author
uses the $\mbox{p}^{*}$ model \cite{Wasserman05} (this has recently been called
the exponential random graph (ERG) model.)
This method reveals whether a network
includes significant configurations (some specific patterns or motifs) or not.
To explain the $\mbox{p}^{*}$ model,
this paper uses a dependence graph model \cite{Koely05}.

\begin{figure}[b]
\begin{center}
\includegraphics[scale=0.8]{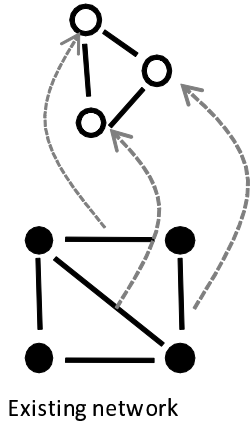}
\caption{Example of dependence graph. Node in dependence graph corresponds to link in actual networks.}
\label{fig:dependence}
\end{center}
\end{figure}

First,
dependence graphs are explained.
They are graphs whose nodes are the links in some network.
Figure \ref{fig:dependence} outlines an example of a dependence graph.
The closed circles at the bottom represent the nodes
in an existing network
and the open circles at the top represent the nodes
corresponding to the links in the lower network.
The links between nodes in a dependence graph
means there is a relationship between the links in an existing network.
Since a dependence graph indicates
what kinds of configurations there are in networks,
the $\mbox{p}^{*}$ model can be understood
by using this.
The example in Figure \ref{fig:dependence} shows
an existing network only has one type of link.
Since this paper treats transaction and joint-patent application
networks, dependence graphs include and distinguish between the two types of links.

It is assumed
in the $\mbox{p}^{*}$ model
that
a network is one entity
where nodes are fixed
and links are probabilistically created.
Network $x$ is a notation
whose links are determined in given nodes.
The probability of $x$ 
in the model
defined as
\begin{equation}
P(x)=\kappa^{-1}\mbox{exp}(\sum_{A\subseteq N_D}\lambda_{A}z_{A}(x))\label{eqn:rand},
\end{equation}
where $\kappa=\sum_{x}\mbox{exp}(\sum_{A\subseteq N_D}\lambda_{A}z_{A}(x))$,
and this is a normalized constant.
$A$ is a sub-class of nodes,
and $N_D$ represents the nodes from the dependence graph.
$x_{ijm}$ is the link from node $i$ to $j$ in network $m$
and $z_{A}(x)=\prod_{x_{ijm}\in A}x_{ijm}$.
Network $m$ is a layer of networks.
$\lambda_{A}$ is a coefficient.
The essence of the $\mbox{p}^{*}$ model is
to find $\lambda_{A}$ so that $P(x)$ is maximized.

There are many redundant configurations 
in Equation (\ref{eqn:rand}) of the $\mbox{p}^{*}$ model
corresponding to $\lambda$.
We have to reduce
the number of $\lambda$s down since this is huge.
Moreover, we can only obtain one existing network.
We can use isomorphism to reduce the number of $\lambda$s.
This means that
identical configurations in the networks are assumed to
have an identical $\lambda$.
The name of the $\mbox{p}^{*}$ model is derived from this operation.

Equation (\ref{eqn:rand}) is updated by considering isomorphism,
and the equation becomes
\begin{equation}
P(x)=\kappa^{-1}\mbox{exp}(\sum_{[A]}\lambda_{[A]}z_{[A]}(x)).
\end{equation}
We have to acquire the $\lambda$s so that
$P(x)$ is maximized for the existing networks.
There are several approaches to
the calculation i.e.,
maximum likelihood estimation
\cite{Wasserman87,Frank93}
and Markov chain Monte Carlo methods \cite{Carrington05}.
These methods have been studied extensively,
but the calculation involved is complicated.
The pseudolikelihood function
\cite{Besag75,Besag77,Strauss90,Pattison99}
is another method that has recently been proposed
and as the calculation involved is simple
it has been used in
this paper.

Three notations should be defined in the
preparation,
\begin{eqnarray}
\mathbf{X}^C_{ijm} & = & \{\mathbf{X}_{klh}:\mathbf{X}_{klh}\in N_D \, \mbox{for all} \, (k,l,h)\neq (i,j,m) \, \mbox{and} \, \mathbf{X}_{ijm} \, \mbox{is undefined} \}\nonumber, \\
\mathbf{x}^+_{ijm} & = & \{x^+_{klh}:x^+_{klh}=x_{klh} \, \mbox{for all} \, (k,l,h)\neq (i,j,m) \, \mbox{and} \, x_{ijm}=1\} \nonumber, \mbox{and} \\
\mathbf{x}^-_{ijm} & = & \{x^-_{klh}:x^-_{klh}=x_{klh} \, \mbox{for all} \, (k,l,h)\neq (i,j,m) \, \mbox{and} \,  x_{ijm}=0\} \nonumber .
\end{eqnarray}
Based on these,
the conditional probability for $X_{ijm}$ with 
a link between nodes $i$ and $j$ in network $m$ is
represented as
\begin{eqnarray}
P(X_{ijm}=1\mid \mathbf{X}^C_{ijm}) & = & \frac{P(\mathbf{X}=\mathbf{x}^+_{ijm})}{P(\mathbf{X}=\mathbf{x}^+_{ijm})+P(\mathbf{X}=\mathbf{x}^-_{ijm})} \nonumber \\
& = & \frac{\exp{\sum_{[A]}{\lambda_{[A]}z_{[A]}(\mathbf{x}^+_{ijm})}}}{\exp{\sum_{[A]}{\lambda_{[A]}z_{[A]}(\mathbf{x}^+_{ijm})}}+\exp{\sum_{[A]}{\lambda_{[A]}z_{[A]}(\mathbf{x}^-_{ijm})}}}. \nonumber
\end{eqnarray}


The pseudolikelihood function method
defines the pseudolikelihood
\begin{small}
\[
PL(\lambda)=\prod_{i\neq j}\prod^{r}_{m=1}P(X_{ijm}=1\mid \mathbf{X}^C_{ijm})^{x_{ijm}}P(X_{ijm}=0\mid \mathbf{X}^C_{ijm})^{1-x_{ijm}}.
\]
\end{small}
The coefficient $\lambda$s that maximize the pseudolikelihood
can be acquired by logistic regression analysis and
this has been proved \cite{Strauss90}.

Deviance is commonly used
to evaluate the importance of models.
That is expressed by $G^2_{PL}$.
We have to establish the reasonable difference in deviance
as to whether a parameter ($\lambda$) contributes to model
networks or not.
If the removal of a single parameter
leads to an increase in the deviance of models that is no more than
$-2n(n-1)r\log(1-\delta)$,
where $n$ is the number of nodes, $r$ is the number
of network layers and $\delta$ is a tunable parameter,
the parameter should be omit.
Following a previous study by Koely and Pattison \cite{Koely05},
$\delta$ was set to 0.001.
This necessary gap was called $\alpha$.

\begin{figure}[b]
\begin{center}
\includegraphics[scale=0.5]{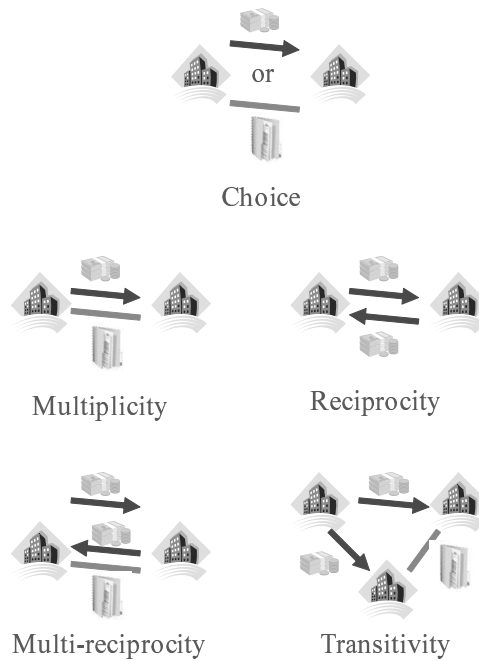}
\caption{Basic and four other configurations.
Possibility of these configurations is investigated in the analysis.}
\label{fig:pattern}
\end{center}
\end{figure}

This paper investigates one base configuration and four configurations,
i.e., 
choice, multiplicity, reciprocity, multi-reciprocity and
transitivity.
They are outlined in Figure \ref{fig:pattern}.
More complicated configurations can be considered.
but they are not necessary.
The reason will be explained later.
In the previous work \cite{Koely05},
both types of links had directions.
However, only one type of links has directions in this paper.
Hence, there are fewer configurations
in terms of variations than in the previous study.

Choice configuration
means the probability that 
exists either link between two nodes.
Therefore, there are two $\lambda$s.
By obtaining deviance for this base model,
we can determine whether
some other model with a single parameter
has a larger difference in deviance to the base model than $\alpha$.

Multiplicity configuration, which is different from that of choice,
means that both types of links between two nodes are represented,
and one $\lambda$ is assigned.
Therefore, this model reveals whether different types of links
simultaneously emerge or not.
The reciprocity configuration
means that there are two-way transaction links between two nodes, and
one $\lambda$ is assigned.
The multi-reciprocity configuration
also has a joint-patent application link.

The transitivity configuration
represents the probability of links
between three nodes.
Other types of configurations between three nodes
can be considered.
However, this paper only presents the results for this configuration
and the reasons will ee explained later.
Suppose that the three nodes are called A, B and C;
this configuration creates a situation where
A and B, and A and C are connected by transaction links.
In addition, B and C are connected by a joint-patent application link.

\begin{table}[h]
\begin{center}
\caption{$G^2_{PL}$ acquired with the $\mbox{p}^{*}$ model (First 20 industries)}
\label{tbl:pstarresult1}
{\small
\begin{tabular}{|ll|rrr|rrrr|}
\hline
\multicolumn{2}{|l|}{Industry} & \multicolumn{1}{l}{Nodes} & \multicolumn{1}{l}{$\alpha$} & \multicolumn{1}{l|}{Choice} & \multicolumn{1}{l}{Multi-} & \multicolumn{1}{l}{Recipro-} & \multicolumn{1}{l}{Multi-} & \multicolumn{1}{l|}{Transi-} \\
 & & & & & \multicolumn{1}{l}{plicity} & \multicolumn{1}{l}{city} & \multicolumn{1}{l}{reciprocity} & \multicolumn{1}{l|}{tivity} \\
\hline
\hline
(01) & Agriculture, forestry & 5 & - & - & - & - & - & - \\
 & and fishery & & & & & & & \\
\hline
(02) & Mining & 72 & 8.9 & 2,282.8 & 2,230.6 & 2,212.6 & 2,269.3 & 2,281.1 \\
 & & & & & * & * & *& \\
\hline
(03) & Foods & 355 & 218.4 & 20,054.6 & 18,453.7 & 19,209.3 & 19,839.7 & 19,928.6 \\
 & & & & & * & * & & \\
\hline
(04) & Textile products & 381 & 251.6 & 2,2823.9 & 21,923.0 & 22,137.5 & 22,744.5 & 22,731.4 \\
 & & & & & * & * & & \\
\hline
(05) & Pulp, paper and & 242 & 101.4 & 10,872.0 & 10,133.7 & 10,547.3 & 10,799.2 & 10,784.2 \\
 & wooden products & & & & * & * & & \\
\hline
(06) & Chemical products & 1,104 & 2,116.4 & 112,351.0 & 107,759.2 & 108,230.3 & 11,562.7 & 110,612.2 \\
 & & & & & * & * & & \\
\hline
(07) & Petroleum and coal & 119 & 24.4 & 5,077.1 & 4,847.1 & 4,981.8 & 5,057.3 & 4,987.4 \\
 & products & & & & *& *& &  *\\
\hline
(08) & Ceramic, stone and & 711 & 877.4 & 58,411.0 & 55,990.4 & 56,533.3 & 58,220.1 & 57,566.6 \\
 & clay products & & & & * & * & &  \\
\hline
(09) & Iron and steel & 636 & 701.9 & 52,940.0 & 50,680.0 & 51,835.0 & 52,926.0 & 52,747.2 \\
 & & & & & * & * & & \\
\hline
(10) & Non-ferrous metals & 521 & 470.9 & 43,581.2 & 42,079.3 & 42,356.9 & 43,390.6 & 43,101.6 \\
 & & & & & * & * & & *\\
\hline
(11) & Metal products & 875 & 1,329.1 & 72,567.8 & 67,557.4 & 70,268.63 & 71,690.4 & 70,830.1 \\
 & & & & & * & * & & * \\
\hline
(12) & General machinery & 1,554 & 4,194.5 & 166,446.0 & 158,330.2 & 162,082.9 & 165,777.5 & 163,636.7 \\
 & & & & & * & * & & \\
\hline
(13) & Electrical machinery & 1,228 & 2,618.8 & 128,806.6 & 122,003.8 & 124,843.3 & 128,618.9 & 127,615.6 \\
 & & & & & * & * & & \\
\hline
(14) & Information and & 499 & 431.9 & 39,865.0 & 37,677.1 & 38,530.7 & 39,763.2 & 39,586.67\\
 & communication & & & & *&* & &  \\
 & machinery & & & & & & &  \\
\hline
(15) & Electrical equipment & 503 & 438.9 & 36,666.1 & 34,513.2 & 35,566.3 & 36,604.7 & 36,253.8 \\
 & & & & & * & * & & \\
\hline
(16) & Transportation & 1,109 & 2,135.7 & 127,329.5 & 117,616.6 & 123,611.1 & 126,817.6 & 125,270.5 \\
 & equipment & & & &* & *& &  \\
\hline
(17) & Precision instruments & 352 & 214.7 & 21,416.5 & 20,427.3 & 20,659.3 & 21,144.1 & 21,120.2 \\
 & & & & & * & * & * & * \\
\hline
(18) & Miscellaneous & 1,096 & 2,085.9 & 93,997.5 & 89,052.7 & 90,924.2 & 93,542.2 & 92,539.2\\
 &  manufacturing & & & &* &* & &  \\
 & products & & & & & & &  \\
\hline
(19) & Construction & 1,021 & 1,810.0 & 129,410.3 & 123,297.5 & 127,415.4 & 129,206.0 & 127,446.1 \\
 & & & & & * & * & & * \\
\hline
(20) & Electricity, gas & 384 & 269.1 & 33,457.3 & 31,895.1 & 32,743.0 & 33,293.9 & 32,659.4 \\
 &  and heat supply & & & & *&* & & * \\
\hline
\end{tabular}
}
\end{center}
\end{table}

\begin{table}[h]
\begin{center}
\caption{$G^2_{PL}$ acquired with the $\mbox{p}^{*}$ model (Last 14 industries)}
\label{tbl:pstarresult2}
{\small
\begin{tabular}{|ll|rrr|rrrr|}
\hline
\multicolumn{2}{|l|}{Industry} & \multicolumn{1}{l}{Nodes} & \multicolumn{1}{l}{$\alpha$} & \multicolumn{1}{l|}{Choice} & \multicolumn{1}{l}{Multi-} & \multicolumn{1}{l}{Recipro-} & \multicolumn{1}{l}{Multi-} & \multicolumn{1}{l|}{Transi-} \\
 & & & & & \multicolumn{1}{l}{plicity} & \multicolumn{1}{l}{city} & \multicolumn{1}{l}{reciprocity} & \multicolumn{1}{l|}{tivity} \\
\hline
\hline
(21) & Water supply & 17 & 0.47 & 263.7 & 260.6 & 261.5 & 257.4 & 257.9 \\
 & and waste & & & & * &*  & * & * \\
 & management services & & & & & & &  \\
\hline
(22) & Commerce & 1,185 & 2,438.6 & 113,923.8 & 107,206.5 & 109,978.5 & 112,926.2 & 111,329.7\\
 & & & & & * & * & & * \\
\hline
(23) & Financial and insurance & 257 & 114.3 & 13,985.6 & 13,766.6 & 13,619.7 & 13,852.2 & 13,890.5 \\ 
 & & & & & * & * & * & \\
\hline
(24) & Real estate & 136 & 31.9 & 5,686.5 & 5,564.9 & 5,572.9 & 5,678.1 & 5,677.9 \\
 & & & & & * & * & & \\
\hline
(25) & Transport & 190 & 62.4 & 9,721.0 & 9,353.9 & 9,539.0 & 9,683.7 & 9,450.1 \\
 & & & & & * & * & & * \\
\hline
(26) & Communication and & 316 & 173.0 & 20,061.1 & 18,471.19 & 19,515.4 & 19,745.45 & 19,146.87 \\
 & broadcasting & & & &* &* & * & * \\
\hline
(27) & Public administration & 0 & - & - & - & - & - & -\\
 & & & & & * & * & & \\
\hline
(28) & Education and research & 55 & 5.2 & 1,827.5 & 1,821.6 & 1,718.7 & 1,674.1 & 1,813.7 \\
 & & & & & * & * & * & * \\
\hline
(29) & Medical service, health & 4 & - & - & - & - & - & -\\
 & and social security & & & & & & & \\
 & and nursing care & & & & & & &  \\
\hline
(30) & Other public services & 0 & - & - & - & - & - & - \\
\hline
(31) & Business services & 639 & 708.6 & 45,513.8 & 43,030.3 & 44,346.8 & 45,182.0 & 44,674.8 \\
 & & & & & * & * & & * \\
\hline
(32) & Personal services & 5 & - & - & - & - & - & - \\
\hline
(33) & Office supplies & 0 & - & - & - & - & - & - \\
\hline
(34) & Activities not elsewhere & 0 & - & - & - & - & - & - \\
 & classified & & & & & & &  \\
\hline
\end{tabular}
}
\end{center}
\end{table}

Before analyzing the multi-layered network with the $\mbox{p}^{*}$ model,
it should be divided into small parts and
reduced by the number of nodes to decrease the calculation space.
Consequently, the multi-layered network
was divided into the industries presented in the previous section.
By dividing the multi-layered network into the networks confined within industries,
we can compare the common characteristics among the networks of industries.

Each node is classified by industry
to divide a multi-layered network
and transaction and joint-patent application links connected
between different industries are omitted.
Then, 34 separated multi-layered networks are created.
The numbers of nodes in a network are reduced
by separating them
but some networks still have too many nodes.

A three-step process is conducted
to futher reduce the number of nodes in networks.
(1) Find a node with a maximum degree in joint-patent application links.
If more than one node has a maximum degree, one of them is randomly chosen.
(2) Select a connected graph by incrementing steps from the node in
joint-patent application links.
(3) Stop incrementing steps if there are no more connected nodes or
the total number of nodes in the connected graph exceeds 1,000.

Tables \ref{tbl:pstarresult1} and \ref{tbl:pstarresult2} list
the results of analyses obtained with the $\mbox{p}^{*}$ model.
From the left of the tables,
each column lists industries, nodes of connected graphs, and $\alpha$.
Furthermore, the $G^2_{PL}$s for choice, multiplicity, reciprocity, multi-reciprocity
and transitivity are given.
If the differences between these $G^2_{PL}$s and that of choice
are greater than $\alpha$, they are listed with asterisks.
That means the model is significant.

As we can see from
Tables \ref{tbl:pstarresult1} and \ref{tbl:pstarresult2},
some industries do not have joint-patent application links,
or have few links.
Industries with fewer than 10 nodes were not analyzed.

Tables \ref{tbl:pstarresult1} and \ref{tbl:pstarresult2}
indicate that multiplicity and reciprocity configurations are
significant in all the analyzed industries.
Multiplicity means
transaction and joint-patent application links
tend to emerge simultaneously,
and reciprocity means
two-way transaction links between two nodes
tend to emerge simultaneously.

Multi-reciprocity and transitivity configurations,
on the other hand,
are significant in some of the analyzed industries.
Multi-reciprocity is a mixture of
multiplicity and reciprocity.
The significance of multi-reciprocity is indicated
in mining, precision instruments,
water supply and waste management services,
financial and insurance, communication and broadcasting,
and education and research.
Before conducting the analyses,
the present author expected that significant configurations
would appear in some specific industries that shared some characteristics.
However, it is difficult to find any common
characteristics in the listed industries.
It is also difficult to find
significance in transitivity in
petroleum and coal products,
non-ferrous metals,
metal products,
transportation equipment,
construction,
electricity, gas and head supply,
water supply and waste management services,
commerce,
transport,
communication and broadcasting,
education and research,
and business services.
I could not find any common characteristics in the industries either.
Hence, we can assume that
other complicated configurations have the same tendency.
This is 
why other complicated configurations were not analyzed.

With the analyses of the $\mbox{p}^{*}$ model explained in this section,
I could not find common characteristics in industries
in the analyses of configurations with three links or three nodes,
and it is assumed that it was not worthwhile to investe other complicated configurations.
However,
the results for multiplicity are important.
All analyzed industries indicated the significance of multiplicity,
and this means that transactions and joint-patent application links 
are closely related.
If there is a relationship between 
transaction and joint-patent application links between two nodes,
the next question is
which type of link precedes the other type.
This wiil be discussed in the next section.

\section{Bayesian networks}

The relationship between transaction and joint-patent application
links was revealed
in Sections \ref{cha:iotable} and \ref{cha:pstar}
at different levels, i.e., industries and firms.
Consequently,
whether either type of link tends to precede the other
should be discussed.
This question corresponds to the old question,
do seeds precede needs, or do needs precede seeds?

The analysis of Bayesian networks is discussed
in this section.
Bayesian networks are completely different from the networks
discussed in this paper thus far and these are methods to
find causality from data.

A Bayesian network is a graphical model
that encodes the joint probability distribution
for a set of random variables.
Some applications can be found in the overview by Lauritzen \cite{Lauritzen03}.
Bayesian networks were first proposed by
Pearl \cite{Pearl09}.
Bayesian networks are used to fulfill three main objectives.
They are to (1) infer unobserved variables,
(2) to achieve parameter learning and
(3) to accomplish structure learning.

Bayesian network can be used to infer unobserved variables
because it is a model for variables and their relationships.
Hence, by using some state of a subset of variables,
Bayesian network updates the relationships and gives the joint probability
distributions.

We often want to know the joint probability distributions among variables.
This is parameter learning.
It is necessary to specify 
the probability distribution 
for each node
conditional on parents.

Structure learning is 
an advanced form of parameter learning.
To acquire an expert's way of thinking
or to dissolve ambiguity in causality in variables,
the network structure and parameters for joint probability distributions
must be learned from the data.
Therefore, structure learning
exactly conforms to the objective of this section.

The method of structure learning is based on 
the idea proposed by Rebane and Pearl \cite{Rebane87}
and there are various kinds of improved methods.
Deal is one methods propose by B\o ttcher and Dethlefsen \cite{Bottcher03}
and can be executed with {\bf R}.

\begin{figure}[b]
\begin{center}
\includegraphics[scale=0.5]{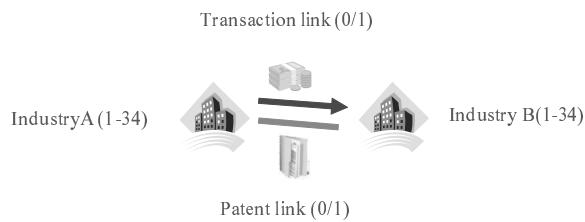}
\caption{Parameters for structure learning of Bayesian network.}
\label{fig:bayesparam}
\end{center}
\end{figure}

Figure \ref{fig:bayesparam} outlines the parameters for structure learning
of Bayesian network.
The parameters are defined for two firms.
As seen in Figure \ref{fig:bayesparam},
Industry A and B parameters are firms' industries
and if there is a joint-patent application link,
the patent link parameter is set to 1, and if not, it is set to 0.
If there is a transaction link from the firm of industry A
to the firm of industry B,
The transaction link parameter is set to 1, and if not, it is set to 0.
Other parameters can be considered
but Bayesian network built on four parameters
is simple and easy to understand.

As the multi-layered network has 975,607 nodes,
there are 975,607$\times$(975,606-1)$/$2 possible pairs.
Since this is too large to calculate,
pairs are chosen in four steps.
(1) Choose one node randomly.
(2) Acquire nodes that can traverse from the start node
in a step with either link.
(3) Consider all pairs from the chosen nodes.
(4) Repeat (1) to (3) until enough pairs are acquired.
This process is necessary because,
as is intuitively understood,
a randomly chosen pair is
seldomly connected by links.
By using the above process,
109,641 pairs are acquired.

\begin{figure}[b]
\begin{center}
\includegraphics[scale=0.2]{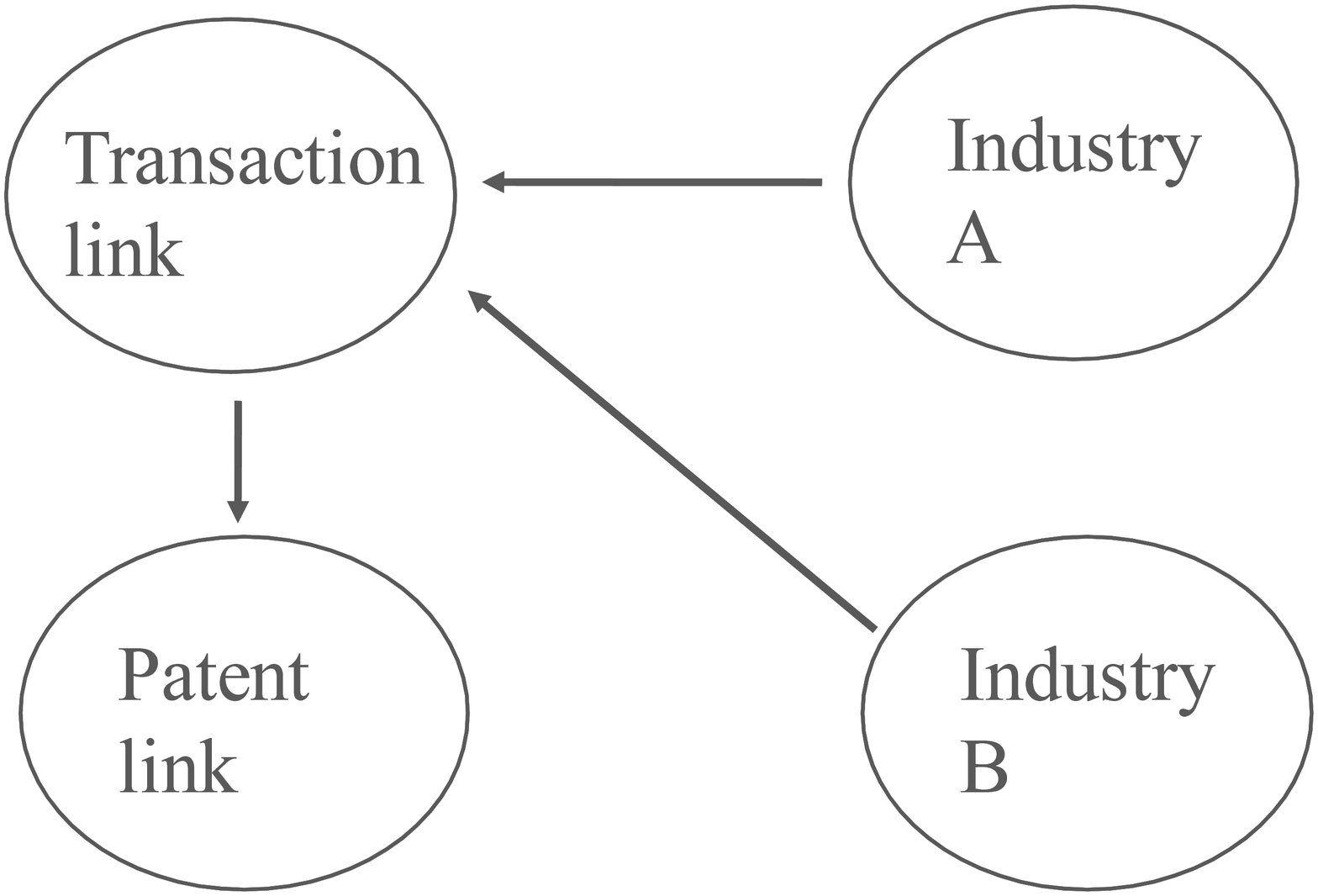}
\caption{Bayesian network acquired by structure learning.}
\label{fig:bayesresult}
\end{center}
\end{figure}

Figure \ref{fig:bayesresult} outlines
the results of structure learning conducted by deal.
It seems that industries A and B precedes the transaction link
and it precedes the patent link,
but this is not true.
From this structure,
we can only learn that
(1) the transaction and patent links are dependent.
(2) The transaction link and industry A, and the transaction link and industry B are dependent.
(3) If the value of the transaction link is given,
the parameters, patent link, and industries A and B are independent.
Since structure learning has limitations,
the direction of dependence between
transaction and joint-patent application links cannot be acquired.
However,
it is important for there to be
many other possible structures
and to choose the structure of the results from them.
An example of the interpretation of 
results is where
the possibility of a joint-patent application link between two firms
only depends on a transaction link and the value of
transaction links between any two firms is assessed in common.
Therefore, we do not have to care about the industries of firms to know
for the value of patent links.

\section{Conclusion}

This paper investigated
the relationship between 
transactions 
and cooperative R\&D.
Two different types of data were used.
The first type was transaction data from Japanese firms.
The network created from the data includes 961,363 firms and
7,808,760 links.
The second type of data was joint-patent applications.
The joint-patent application network included
54,197 nodes and
154,205 links.
These two networks were merged into one network,
called a multi-layered network.
The network included
975,607 nodes and both types of links.

The first anaysis was based on input-output tables
and three different tables were compared.
The correleation coefficient between three pairs of tables
were 0.31 (total amount of money and joint-patent applications),
0.45 (transactions and joint-patent applications)
and 
0.05 (transactions and the total amount of money).
Transactions were found to be more strongly tied to joint-patent applications
than the total amount of money.
This means 
that the total amount of money can
be used to track the 
scientific and technological connections to some degree,
but transactions are 
more preferable to reach the objective.
The total amount of money and transactions
have few relationships
and these are probably connected to the joint-patent applications
in different mechanisms.

The second analysis was conducted based on 
the $\mbox{p}^{*}$ model.
This revealed whether a network
included significant configurations or not.
The configurations of
choice, multiplicity, reciprocity, multi-reciprocity and
transitivity were evaluated.
In this analysis, the multi-layered network
was divided into industries.
Multiplicity and reciprocity configurations were
significant in all the analyzed industries.
Multiplicity meant
transactions and joint-patent application links
tended to emerge simultaneously.
The results for multiplicity 
meant that transactions and joint-patent application links 
were closely related.
Reciprocity meant
two-way transaction links between two nodes
tended to emerge simultaneously.
Multi-reciprocity and transitivity configurations were
significant in some of the analyzed industries.
It was difficult to find any common
characteristics in the industries.

Bayesian networks were used
in the third analysis.
Bayesian networks are methods of finding causality from data
and one usage is structure learning that can be executed by deal.
The parameters were defined on two firms and they were
Industries A and B, patent links and 
transaction links.
By using the process of reducing pairs,
109,641 pairs were acquired.
From the learned structure,
we knew that
(1) transaction links and patent links were dependent.
(2) transaction links and Industry A and transaction links and Industry B were dependent.
(3) If the value of a transaction link is given,
the parameters of patent links and Industries A and B are independent.

\section{Acknowledgements}

This work was supported by KAKENHI (20730321).

\bibliographystyle{unsrt}
\bibliography{paper}

\end{document}